\documentclass{article}

\usepackage{url}
\usepackage{titlesec}
\usepackage{authblk}
\usepackage[outdir=./]{epstopdf}
\usepackage{graphicx,changepage}

\begin{document}



\title{matLeap: A fast adaptive Matlab-ready tau-leaping
implementation suitable for Bayesian inference}
\author{Justin Feigelman \thanks{justin.feigelman@imsb.biol.ethz.ch}}
\author{Stefan Ganscha \thanks{stefan.ganscha@imsb.biol.ethz.ch}}
\author{Manfred Claassen
\thanks{manfred.claassen@imsb.biol.ethz.ch; Corresponding author} }
\affil{Institute of Molecular Systems Biology, ETH Z\"urich, Z\"urich, 8093, Switzerland}

%

\maketitle

\abstract{\textbf{Background:} Species abundance distributions in chemical
reaction network models cannot usually be computed analytically.
Instead, stochastic simulation algorithms allow sample from the the system configuration. 
Although many algorithms have been described, no fast implementation has been
provided for $\tau$-leaping which i) is Matlab-compatible, ii) adaptively
alternates between SSA, implicit and explicit $\tau$-leaping, and iii) provides
summary statistics necessary for Bayesian inference.
\\
\textbf{Results:} We provide a Matlab-compatible implementation of the adaptive
explicit-implicit $\tau$-leaping algorithm to address the above-mentioned
deficits. matLeap provides equal or substantially faster results compared to 
two widely used simulation packages while maintaining accuracy. Lastly, matLeap yields
summary statistics of the stochastic process unavailable with other methods, which are
indispensable for Bayesian inference.\\
\textbf{Conclusions:} matLeap addresses shortcomings in existing
Matlab\--compatible stochastic simulation software, providing significant
speedups and summary statistics that are especially useful for researchers
utilizing particle-filter based methods for Bayesian inference. Code is available for
download at \url{https://github.com/claassengroup/matLeap}. \\
\textbf{Contact:} \url{justin.feigelman@imsb.biol.ethz.ch}\\
}

\section{Background}

Chemical reaction networks (CRNs) provide a quantitative description of the
probabilistic evolution of systems of interacting molecules, and are frequently
used to model biological systems such as gene regulatory or signaling networks.
However, CRNs are not generally solvable in closed form. One may
instead produce sample trajectories of the stochastic process dependent on 
initial conditions, reaction stoichiometry and parameters using the
Stochastic Simulation Algorithm (SSA) \cite{Gillespie:1977uf} and related
algorithms. However, SSA is prohibitively expensive for systems
with disparity in relevant time scales \cite{Gillespie:2007uq}.
Algorithms exist for exploiting the difference in time scales including
$\tau$-leaping algorithm and variants \cite{Gillespie:2001vc}, which reduce
computational effort by approximating the Markov jump process by a Poisson process over intervals for which reaction probabilities are approximately constant. 
The jump intervals are chosen to bound the expected
change (and variance) in reaction probabilities, providing a tunable control for
accuracy \cite{Cao:2006tb}.

Several implementations of stochastic simulation algorithms are
available, often combined with extensive graphical environments for creating,
simulating and analyzing the results, see e.g.\ Systems Biology Toolbox
\cite{Anonymous:EJ3UZCnd}, or Matlab SimBiology. However, they are
limited to either explicit or implicit $\tau$-leaping and do not adaptively
change in response to dynamic system stiffness. To address this gap we developed matLeap, a fast
C++ based implementation of the adaptive explicit-implicit 
$\tau$-leaping algorithm \cite{Cao:2007ki}. matLeap adapts
between $\tau$-leaping algorithms with the stiffness of the system, and ensures
non-negativity of species by using critical reactions and switching to SSA when
required. Unlike most available packages, we focus on a minimalistic
interface-free implementation which integrates directly with Matlab, using models provided as SBML files and a few tuning parameters to perform simulations.

For Bayesian parameter inference of CRNs, one is often
interested in inferring the posterior distribution of the reaction constant $\theta_i$ of reaction $i$.
Assuming mass action kinetics, the reaction propensity at time $t$ is
given by $a_i(X_t) = \theta_i~g_i(X_t)$, where $g_i(X_t)$ is a function of the reaction
educt copy numbers and $X_t$ is the state of the system. Inference is simplified
if one assumes gamma-distributed reaction constants, i.e.\ $\pi_i(\theta_i) =
\Gamma(\theta_i; \alpha_i,\beta_i)$ with hyperparameters $\alpha_i,\beta_i$. In
this case the posterior distribution for a trajectory with $r_i$ firings of
reaction $i$ and integral $G_i = \int g_i(X_s) ds$ of the function $g_i$ is given by $\Gamma(\theta_i; \alpha_i
+ r_i, \beta_i + G_i)$ \cite{Golightly:2006hp}.  
matLeap provides the summary statistics, $r_i$ and
$G_i$ for each reaction $i$; $G_i$ is
computed using trapezoidal approximation when $\tau$-leaping and exactly when
performing SSA. Lastly, the implicit $\tau$-leaping
algorithm requires the inverse Jacobian of the reaction propensities. For small
systems this can be computed symbolically in Matlab and supplied to
matLeap for additional speedup; otherwise it is estimated numerically.

\section{Implementation}
matLeap implements the adaptive explicit-implicit $\tau$-leaping
algorithm \cite{Cao:2007ki}, switching to implicit
$\tau$-leaping if the computed leap size is much greater when excluding
reaction pairs in equilibrium. Non-negativity of species is ensured
using critical reactions. We slightly modified the algorithm to switch to SSA
if the waiting to the next critical reaction is comparable to 
the waiting time to the next reaction preventing critical, very fast reactions
from forcing the system to perform $\tau$-leaping over very small intervals.

matLeap is implemented in C/C++ using the Eigen 
\cite{eigenweb} and Boost (http://boost.org) libraries.
It generates a mex-file that can be called with
variable initial conditions and model parameters; control parameters
determine the accuracy of the $\tau$-leaping approximation, and behavior with
respect to critical species (see Supplemental Information). 
Models are specified
using SBML and loaded using libSBML \cite{Bornstein:2008wg}. 
The Matlab symbolic computing and compiler toolboxes are required.

The mex file can be called with a matrix of initial conditions for all
species, and/or a matrix of parameter values to be used which is especially
useful for parameter inference settings. We note that this is in contrast to
StochKit and SimBiology which require the user to create new configuration files or structures, respectively, before rerunning. 

\section{Results}
\begin{figure}[htp]
\begin{center}
  \includegraphics[width=\textwidth]{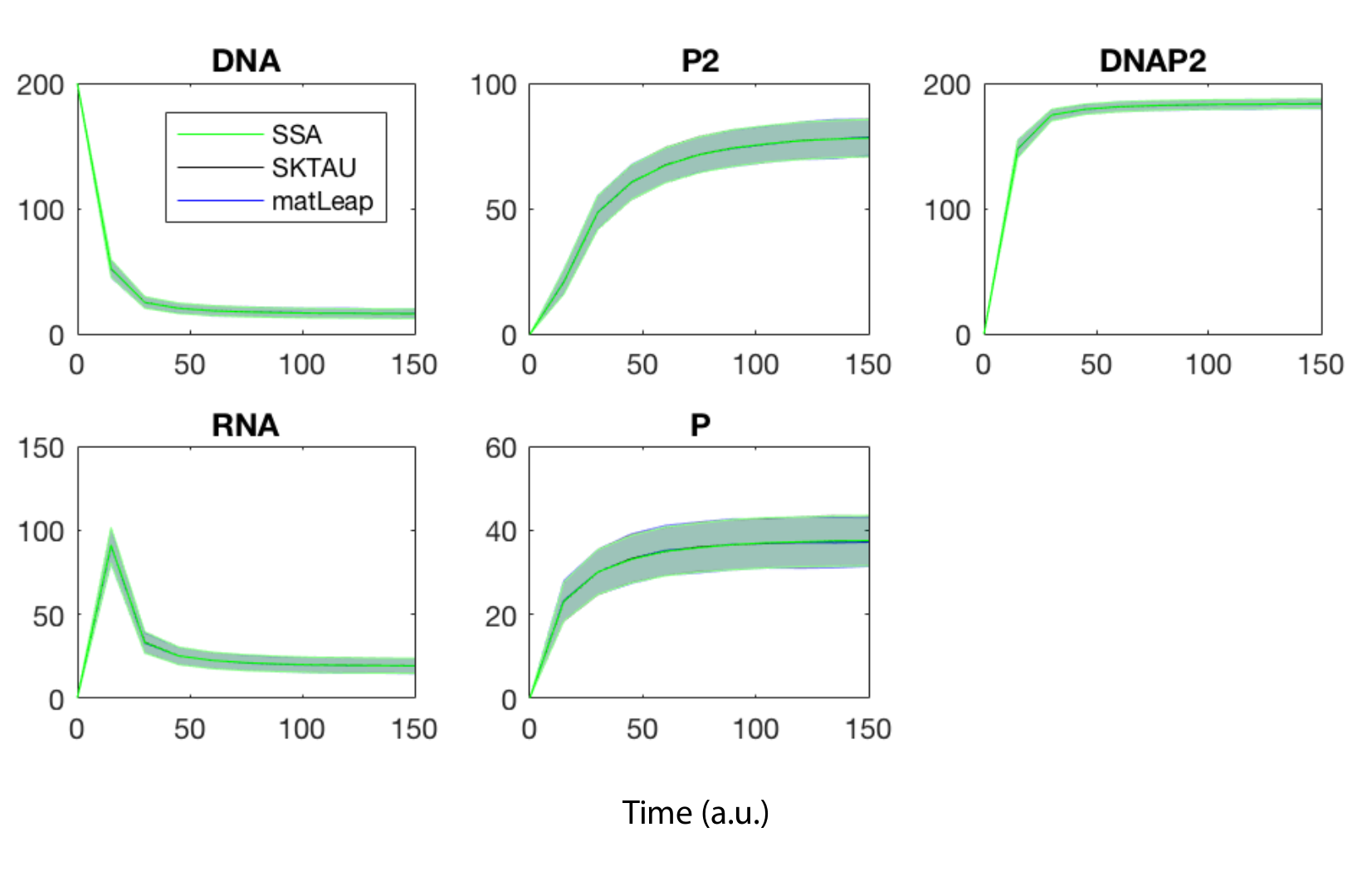}
  \caption[labelInTOC]{Comparison of SSA, $\tau$-leaping using StochKit, and
  $\tau$-leaping using matLeap for the prokaryotic auto-regulatory gene network model
\cite{Wang:2010ie}}
  \label{fig:Wang2010}
\end{center}
\end{figure}

\begin{figure}[htp]
\begin{center}
  \includegraphics[width=\textwidth]{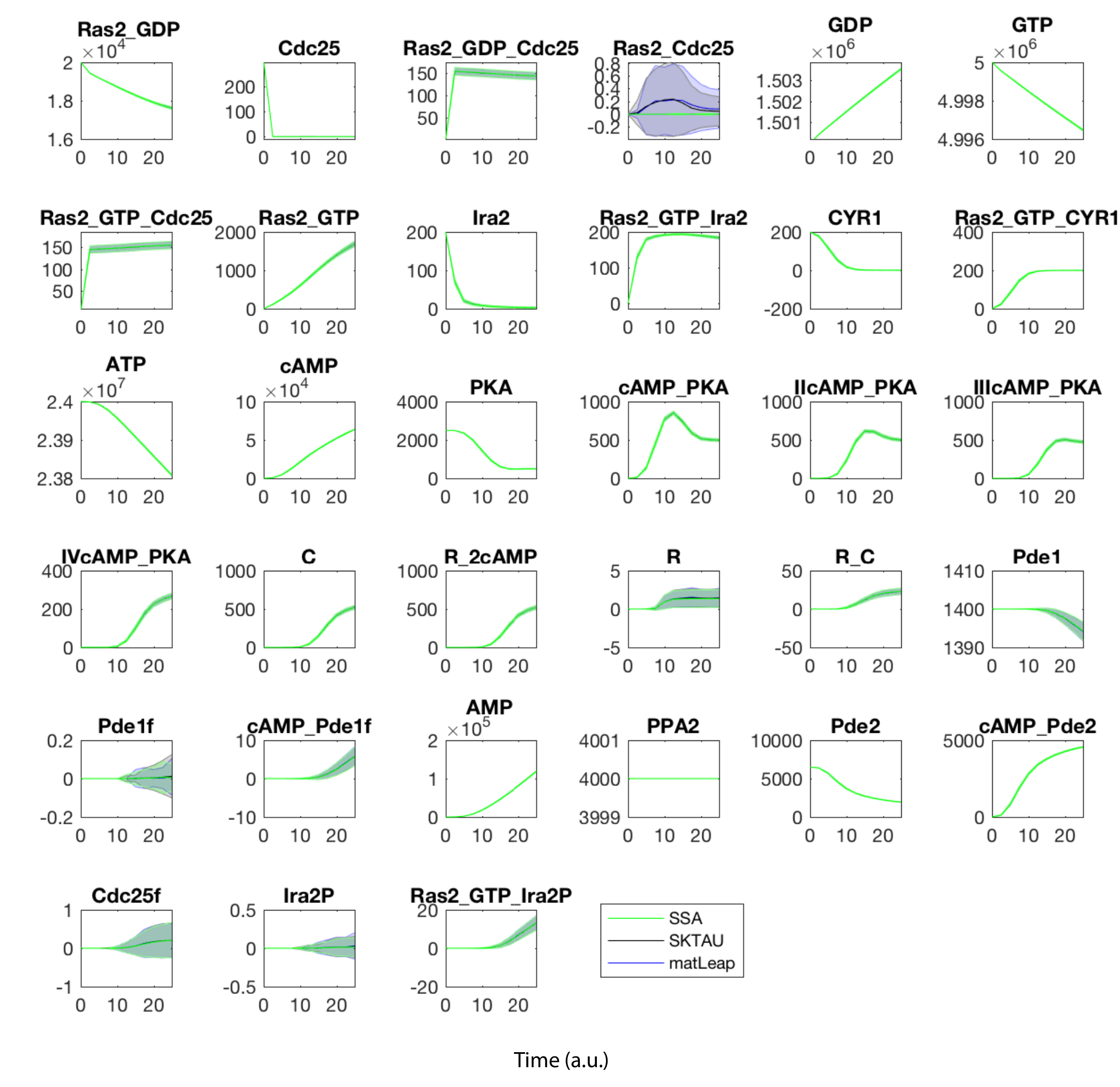}
  \caption[Besozzi2012]{Comparison of SSA, $\tau$-leaping using StochKit, and
  $\tau$-leaping using matLeap for the Ras/cAMP/PKA pathway \cite{Besozzi:dd}}
  \label{fig:Besozzi2012}
\end{center}
\end{figure}

\begin{figure}[htp]
\begin{center}
  \includegraphics[width=\textwidth]{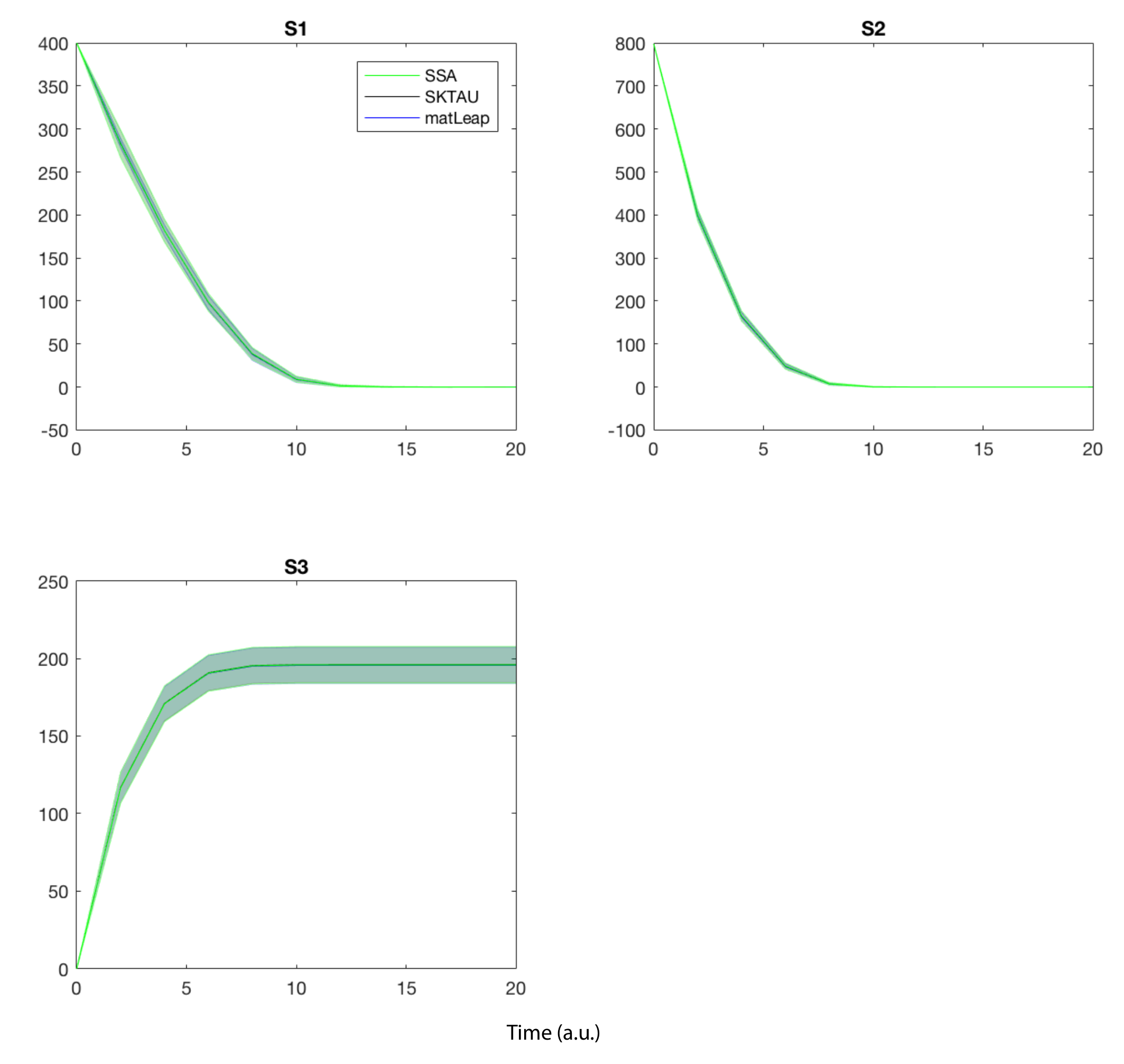}
  \caption[Rathinam2003]{Comparison of SSA, $\tau$-leaping using StochKit, and
  $\tau$-leaping using matLeap for the stiff decaying-dimerizing reaction set
  \cite{Rathinam:2003jea}}
  \label{fig:Rathinam2003}
\end{center}
\end{figure}

We compared matLeap against two frameworks, StochKit 2.0
\cite{Sanft:2011cj} and the Matlab SimBiology toolbox, for three models: the prokaryotic auto-regulatory gene network model
\cite{Wang:2010ie} (Figure \ref{fig:Wang2010}), the Ras/cAMP/PKA pathway in
\textit{S. cerevisiae} \cite{Besozzi:dd} (Figure \ref{fig:Besozzi2012}), and a
stiff decaying-dimerizing reaction set \cite{Rathinam:2003jea} (Figure \ref{fig:Rathinam2003}).

In each case we compute the accuracy, given by the Kullback-Leibler divergence, averaged over all time points and species, 
with respect to exact SSA (Table \ref{tab:Accuracy}). We used $10^4$ stochastic
simulations and 5 replicates, see Supplemental Information for comparison of
simulated trajectories.
For the $\tau$-leaping methods, we set the control parameter $\epsilon$ to 0.03.
StochKit $\tau$-leaping determines automatically at run-time the appropriate
(explicit or implicit) algorithm.

matLeap is as accurate as StochKit $\tau$-leaping except for
the dimerization model, where it underestimates the variance of two fast
species (c.f. Figure 3).
This is a known limitation of the implicit $\tau$-leaping algorithm
\cite{Rathinam:2003jea}. However, matLeap can be run in fully explicit
$\tau$-leaping or SSA-mode if higher accuracy for the variances of fast species
is needed. In each case matLeap runs at least as fast as the faster of StochKit
SSA or $\tau$-leaping; in some cases it runs nearly 100 times as fast.
For the dimerization model using the symbolic Jacobian provides an
additional $\approx 14\%$ speedup.
Furthermore, matLeap runs between approximately 6 and 350 times faster than
Matlab SimBiology. Due to the excessive run time, we did not evaluate the accuracy of Matlab SimBiology. We note that
StochKit necessarily writes all results to disk (as plain text) which
contributes to its run time. Also, SimBiology saves the complete reaction path
of all species, which significantly slows its performance and greatly increases
memory consumption. 

\begin{table}
\begin{tabular}{|p{2cm}|p{1.25cm}|p{1.25cm}|p{1.25cm}|p{1.25cm}|p{1.25cm}|p{1.25cm}|}
\hline
& \multicolumn{2}{|p{1cm}|}{Auto-regulation} &
\multicolumn{2}{p{2cm}|}{Ras/cAMP/PKA} & \multicolumn{2}{p{2.5cm}|}{Stiff
dimerizing/decaying}
\\
\hline
Method 					&  KL-div. 				& Run time 	& KL-div. 				& Run time & KL-div. 				& Run time \\
\hline                                                                                             
StochKit SSA			&  $0.0034 \pm 0.0013$ 	& 0.0110 	& $0.0021 \pm 0.0014$ 	& 0.1464 & $0.0023 \pm 0.0017$ 	& 0.2605 \\	
StochKit $\tau$-leap	&  $0.0036 \pm 0.0012$ 	& 0.1803 	& $0.0051 \pm 0.0172$ 
& 0.1891 & $0.0023 \pm 0.0018$  	& 0.4018 \\
matLeap 				&  $0.0037 \pm 0.0012$ 	& 0.0046 	& $0.0057 \pm 0.0187$ 	& 0.1472 & $0.0408 \pm 0.1213$ 	& 0.0042 \\	
matLeap (symbolic) 		&  $0.0038 \pm 0.0013$ 	& 0.0046 	& n.a. 					& n.a.   & $0.0414 \pm 0.1220$ 	& 0.0037 \\	
SimBiology 				&  n.a. 				& 0.0419 	& n.a. 					& 0.8574 & n.a. 					& 1.4818 \\
\hline
\end{tabular}
\caption{Average (std.) Kullback-Leibler divergence and run-time comparison of
matLeap and StochKit evaluated for $\epsilon=0.03$ ($10^4$ simulations for
StochKit and matLeap, 20 for SimBiology). Run times reported as average per
simulation (all standard deviations $<$ 4\%, n=5).}
\label{tab:Accuracy}
\end{table}

\section{Discussion}
Stochastic simulation is a mature field with many exact and
approximate solvers available. However, few of the existing methods are
available for direct use with Matlab. Comprehensive packages such as
Matlab SimBiology can prove difficult to configure, prohibitively slow, and do
not adaptively switch to accommodate varying problem stiffness. We take an
alternative approach providing a simple and very fast adaptive $\tau$-leaping
solver aimed at practitioners. matLeap is at least as fast as current widely
used implementations, while also uniquely providing the summary statistics $r_i$
and $G_i$, introduced above, which are very valuable for Bayesian inference in
chemical reaction networks. An example of multicore parallelization is
included in the Supplementary Information.

\section{Declarations}
\subsection*{Availability of data and material}
The datasets generated during and/or analysed during the current study are
available at \url{https://github.com/claassengroup/matLeap}, as is the matLeap
package.
\subsection*{Competing interests}
The authors declare that they have no competing interests.

\subsection*{Funding}
This work was supported by the RTD HDL-X grant from SystemsX.ch.

\subsection*{Authors' contributions}
JF conceived of and implemented the numerical method. SG performed testing and
implemented the Matlab package. JF wrote the manuscript. MC provided support and
supervision.

\bibliographystyle{plain}
\bibliography{main_short}

\begin{thebibliography}{10}

\bibitem{Besozzi:dd}
Daniela Besozzi, Paolo Cazzaniga, Dario Pescini, Giancarlo Mauri, Sonia
  Colombo, and Enzo Martegani.
\newblock {The role of feedback control mechanisms on the establishment of
  oscillatory regimes in the Ras/cAMP/PKA pathway in S. cerevisiae}.
\newblock {\em EURASIP journal on bioinformatics {\&} systems biology},
  2012(1):1, July 2012.

\bibitem{Bornstein:2008wg}
Benjamin~J Bornstein, Sarah~M Keating, Akiya Jouraku, and Michael Hucka.
\newblock {LibSBML: an API Library for SBML}.
\newblock {\em Bioinformatics}, 24(6):880--881, March 2008.

\bibitem{Cao:2006tb}
Yang Cao, Daniel~Thomas Gillespie, and Linda~R Petzold.
\newblock {Efficient step size selection for the tau-leaping simulation
  method}.
\newblock {\em The Journal of chemical physics}, 124(4):044109, January 2006.

\bibitem{Cao:2007ki}
Yang Cao, Daniel~Thomas Gillespie, and Linda~R Petzold.
\newblock {Adaptive explicit-implicit tau-leaping method with automatic tau
  selection}.
\newblock {\em The Journal of chemical physics}, 126(22):224101, 2007.

\bibitem{Gillespie:1977uf}
Daniel~Thomas Gillespie.
\newblock {Exact stochastic simulation of coupled chemical reactions }.
\newblock {\em The journal of physical chemistry}, 81(25), 1977.

\bibitem{Gillespie:2001vc}
Daniel~Thomas Gillespie.
\newblock {Approximate accelerated stochastic simulation of chemically reacting
  systems}.
\newblock {\em The Journal of chemical physics}, 115(4):1716--1733, 2001.

\bibitem{Gillespie:2007uq}
Daniel~Thomas Gillespie.
\newblock {Stochastic simulation of chemical kinetics}.
\newblock {\em Annu Rev Phys Chem}, 2007.

\bibitem{Golightly:2006hp}
Andrew Golightly and Darren~J Wilkinson.
\newblock {Bayesian sequential inference for stochastic kinetic biochemical
  network models}.
\newblock {\em Journal of Computational Biology}, 13(3):838--851 (electronic),
  2006.

\bibitem{eigenweb}
Ga~e~l Guennebaud, Beno i~t Jacob, and {others}.
\newblock {Eigen v3}.
\newblock Technical report, 2010.

\bibitem{Rathinam:2003jea}
Muruhan Rathinam, L~R Petzold, and Y~Cao.
\newblock {Stiffness in stochastic chemically reacting systems: The implicit
  tau-leaping method}.
\newblock {\em The Journal of chemical physics}, 119(24):12784, 2003.

\bibitem{Sanft:2011cj}
Kevin~R Sanft, Sheng Wu, Min Roh, Jin Fu, Rone~Kwei Lim, and Linda~R Petzold.
\newblock {StochKit2: software for discrete stochastic simulation of
  biochemical systems with events.}
\newblock {\em Bioinformatics}, 27(17):2457--2458, September 2011.

\bibitem{Anonymous:EJ3UZCnd}
H~Schmidt and M~Jirstrand.
\newblock {Systems Biology Toolbox for MATLAB: a computational platform for
  research in systems biology}.
\newblock {\em Bioinformatics}, 22(4):514--515, 2006.

\bibitem{Wang:2010ie}
Yuanfeng Wang, Scott Christley, Eric Mjolsness, and Xiaohui Xie.
\newblock {Parameter inference for discretely observed stochastic kinetic
  models using stochastic gradient descent.}
\newblock {\em BMC systems biology}, 4(1):99, 2010.

\end{thebibliography}
\end{document}